# Molecular dynamics simulations of cumulative helium bombardments on tungsten surfaces

**Min Li, Jiechao Cui, Jun Wang, Qing Hou**[*]

*Key Lab for Radiation Physics and Technology, Institute of Nuclear Science and Technology,*

*Sichuan University, Chengdu 610061, China*

**Abstract**

Molecular dynamics (MD) simulations were performed to study the cumulative bombardments of low-energy (60-200 eV) helium (He) atoms on tungsten (W) surfaces. The behaviour of He and the response of the W surface were investigated. The He incident energy and tungsten temperature play important roles on the formation and growth of He clusters. The temperature can promote the coalescence of He clusters and increase the size of the He clusters. A stratification phenomenon of the He depth distribution has been observed. The rupture of the He clusters and, at the same time, the escape and re-deposition of the W atoms have been observed. During the formation of He clusters, the interstitial W atoms are produced and evolve into bundles of <111> crowdions, which would be constrained around the He clusters for a long time. However, they will finally move onto the top surface along the <111> direction, which results in stacking the W atoms on the surface. The complex combination effects of the He clusters and the interstitial atoms result in the growth of the surfaces.

[*] Corresponding author. Tel.: +86 28 85412104; fax: +86 28 85410252.

*E-mail address*: qhou@scu.edu.cn (Q. Hou)

## 1. Introduction

Tungsten (W) has been chosen as a promising candidate for the plasma-facing material (PFM) in controlled nuclear fusion reactors (CNFRs) [1-2], especially for the divertors, where the PFM will face irradiation of high-flux particles and intense heat flow [3]. For example, in the International Thermonuclear Experimental Reactor (ITER) [4], the estimated flux of the helium with energy ranging from 20 eV to 200 eV is $(0.1-1)\times 10^{23}$ ions m$^{-2}$ s$^{-1}$, and the temperature of the PFMs is 300 K-3000 K [5]. For the demo or commercial CNFRs in the future, the particle flux and heat flow would be more intense. Thus, interactions between low energy and high fluence helium beams and tungsten surfaces at various temperatures have been the focus of many experimental studies [6-9]. There has been evidence that the condition of high fluence, high temperature and low energy can invoke complex physical processes that are not well understood [10]. To understand the underlying atomistic processes, theoretical studies are needed. Because a large number of atoms are involved in the processes and many-body effects therein act, molecular dynamics is the most suitable method of simulating at an atomistic level the bombardment of low-energy atoms on surfaces.

Using the molecular dynamics, we have studied in a previous paper the damage to tungsten surfaces by the bombardment of low-energy He atoms in which the bombardments were independent of each other [11], a situation that approximates low fluence irradiations. In the present study, we focus on the cumulative bombardment of He atoms on a tungsten surface where a projectile and the defects induced by it will interact with the defects that are produced by previous bombardments. Large-scale molecular dynamics simulations were performed to gain a comprehensive picture of the whole process of the slowing-down of the incident He atoms and the migration, coalescence, dissociation and release of helium and interstitial clusters. After an outline of the simulation method is given in section 2, in section 3.1, the effects of the incident energy, temperature and preexisting vacancies on the nucleation and the release of He will be presented. In section 3.2, the damage of the tungsten surface will be discussed.

## 2. Simulation methods

The molecular dynamics simulation of cumulative bombardments of atoms on surfaces requires computations that are very time consuming. To improve the computational efficiency, a molecular dynamics package using graphics processing units (GPUs) for parallel computing was used [12]. With respect to the physical models, a Finnis-Sinclair-type potential obtained by Ackland et al. [13] was adopted to describe the interactions between the W atoms. An exponential-six potential was employed to describe He-He interactions [14-15]. For the interactions between the He and W atoms, a pair-wise potential, which had been reported elsewhere in detail [16], was used. For the energy exchange between an energetic atom and electrons, we adopted an electron-phonon coupling model [17], in which the temperature of the electrons remained at a given temperature.

The incident energy of the He atoms was selected to be in the range of 60 eV to

200 eV, and the incident direction was set as normal to the tungsten (001) surface. Before initiating the bombardment process, substrate boxes were thermalized and relaxed to thermal equilibrium at the temperature under consideration. The box sizes were chosen to be large enough to eliminate the artifacts that could be introduced by the periodic conditions applied in the x- and y-directions. Because the time for the slowing down of a projectile is approximately 1 ps, for each bombardment, the box was released for 8–10 ps (depending on the projectile energy), which is long enough to bring the system to equilibrium and to avoid continuous heating-up of the system. This simulation box was then used for the next He bombardment. We performed 2000-10000 He impacts for each considered combination of incident energy $E_i$ and substrate temperature $T_s$. In Table 1, the box sizes and the corresponding He fluences that were reached in our simulations are given for different incident energies $E_i$.

## 3. Results and discussion

### 3.1 Nucleation and release of He

Experimental results have shown that bombardments of low-energy He could produce a large amount of bubbles. For example, D. Nishijima et al. found that holes are formed with incident He energy down to 10 eV on a tungsten surface above 2000 K [6]. In this section, we analyse the physical parameters that could affect the nucleation and growth of He clusters on tungsten surfaces at high temperatures.

*3.1.1 Incident energy*

Fig. 1 shows the number of escaping He atoms from tungsten surfaces at 1500 K as a function of the number of injected He atoms for the incident energy $E_i$ =60 eV, 100 eV and 200 eV, respectively. The data was acquired every 50 He bombardments. For $E_i$ =60 eV, the He atoms trapped in the tungsten substrate are few (approximately twenty), and only one small He cluster of 10 He atoms was observed after 2000 bombardments. With increasing incident energy, the rate of He retention increases. There are over one hundred (123, for example) and six hundred (646, for example) He atoms in the tungsten substrate after 2000 bombardments for $E_i$ =100 eV and $E_i$ =200 eV, respectively, and formations of He clusters were observed. It should be noted that the corresponding He fluences at this point (2000 bombardments) are $0.5 \times 10^{20}$ m$^{-2}$ for $E_i$ = 60 eV and 100 eV, and $0.22 \times 10^{20}$ m$^{-2}$ for $E_i$ = 200 eV. The fluences are much lower than in the experiments in [6]. To observe the effect of cumulative bombardments for $E_i$ =60 eV, a much higher He fluence is required. However, because of the high reflection coefficient at low-incident energy [18], most of the bombardments would be 'wasted'. Thus, we will consider in the following only the cases of $E_i$ =100 eV and $E_i$ =200 eV, for the reason of computational efficiency.

Fig. 2 exhibits the comparison of the configurations of $E_i$ =100 eV and $E_i$ =200 eV at the point that the retention dose per area of the He atoms achieves the same level in both cases. The substrate temperature in both cases is 1500 K. The number of He atoms in the box of $63.3 \times 63.3 \times 94.9$ Å$^3$ is 238 after 4000 bombardments for $E_i$ =100 eV, and the number of He atoms is 535 in the box of size $94.9 \times 94.9 \times 94.9$ Å$^3$ after 1669 He bombardments for $E_i$ =200 eV. In the case of $E_i$ =100 eV, some He atoms and small clusters are observed diffusing into deeper places, while most single He atoms and He clusters are distributed near the surface; to be more exact, they are three or four monolayers below the surface. This depth is coincident with the maximum distribution depth of approximately 6 Å, which is obtained for noncumulative bombardments of He on tungsten surfaces with $E_i$ =100 eV [18]. Thus, at this depth, the He density is high and easily induces the formation of He clusters. In the case of $E_i$ =200 eV, in which the He projectiles can penetrate deeper into the substrate than in the case of $E_i$ =100 eV, stratification of the He distribution is observed. Large clusters form around the maximum distribution depth 20-25 Å of noncumulative bombardments. Single He atoms and small clusters are observed accumulating below the surface for three or four monolayers. In our previous studies [11], it was shown that a He atom that backscattered from deeper locations can induce a replacement sequence along the <111> direction, and the last W atom in the sequence is displaced to a stacking site on the top monolayer. This phenomenon occurs most likely when the first W atom in the sequence is three or four monolayers below the surface, because the threshold energy for generating the replacement sequence is the lowest. The substitutional He atom hardly migrates away and then attracts other He atoms to form a cluster. However, it will be shown later that the clusters that are close to the surface are easy to rupture, and it is unlikely that large clusters appear there for both the cases $E_i$ =100 eV and $E_i$ =200 eV.

*3.1.2 Surface temperature*

To study the influence of the substrate temperature $T_s$ on the He cluster nucleation and growth, we ran simulations for 200 eV He atoms bombarding on tungsten substrates at different temperatures. In Fig. 3, we compare the configurations of $T_s = 300$ K and $T_s = 1500$ K at the point that the amount of retention of He atoms is 802 for both cases. As will be shown in Fig. 4, in the initial stage of cumulative bombardments, the He atoms can escape from the surface more easily at high temperature than at low temperature. It took 1498 bombardments for the number of retention He atoms to achieve 802 in the case of $T_s = 300$ K, while the number of bombardments was 2438 in the case of $T_s = 1500$ K. For $T_s = 300$ K, the stratification phenomena of the He depth distribution mentioned above is also observed. The space distribution of He in the case of $T_s = 1500$ K is less dispersed, and the sizes of the He clusters on average are larger than what is observed in the case of $T_s = 300$ K. This observation indicates that increasing the substrate temperature accelerates the migration of He interstitials and, at the same time, promotes the collision and coalescence between them and leads to the formation of large He clusters, especially in the area in which the incident He atoms are slowing down. The large clusters hardly migrate and thus form a "cluster layer" with a few small clusters that are migrating to deeper locations.

Fig. 4 displays an overall view of the escaping rate of He atoms against the number of bombardments. The He atoms escape from the surface through diffusion and rupture of small He clusters near the surface. In the initial stage of the cumulative bombardments, large He clusters are still not formed. Single He atoms and small clusters diffuse faster in the case of $T_s = 1500$ K than in the case of $T_s = 300$ K. Thus, in this stage, the escaping rate of He is higher for $T_s$ = 1500 K than for $T_s$ = 300 K. The He clusters grow along with the accumulation of the bombardments. For $T_s$ = 1500 K, the increasing probability for large He clusters absorbing He atoms tends to prevent the accumulation of He atoms near the surface and leads to a decreasing tendency in the escape rate of the He atoms. In contrast, in the case of $T_s$ = 300 K, where the cluster sizes are smaller, the number of He atoms that accumulate near the surface increases with the accumulation of the bombardments, and the escaping rate of the He atoms exhibits an increasing tendency. Obviously, by the fluence level reached by our simulations, the rate of the He atoms accumulating near the surface and the rate of implementing the He atoms is not in balance.

*3.1.3 Rupture of He clusters*

He atoms escape the surface through temperature-driven diffusion and also through ruptures of the He clusters. In the cumulative simulation process, we observed clusters rupture frequently. A bombardment of a He atom could trigger the rupture of a preexisting near-surface He cluster that is in the neighbouring region of but not necessarily closest to the bombardment entry point. Fig. 5 demonstrates an example of this phenomenon. A He atom of 100 eV bombarded the tungsten surface at a point that was approximately 4 lattice lengths away from the cluster marked 'A'. It is likely that the bombardment introduced changes in the local temperature and the stress of the cluster 'A'. The rupture of cluster 'A' occurred at approximately 0.7 ps after the bombardment. The rupture process is similar to that described in ref. [19] for the helium bubble rupture on titanium surfaces. In the rupture process, the W atoms nearby became disordered and some, but only a few of them, escaped. However, the disorder disappeared quickly, and the tungsten surface structure restored to a nearly perfect crystal structure after the release of the He atoms in cluster 'A'. In this demonstration example, all of the He atoms in cluster 'A' escaped. We also observed situations in which only part of the atoms in a He cluster near the surface dissociated from the cluster and escaped. The He cluster that was left behind absorbed other He atoms and grew, and then, it dissociated again. Thus, large He clusters hardly appear near the surface, as is shown in Fig. 2-3.

**3.2 Damage to the tungsten substrate**

In addition to the ejection of W atoms during the rupture of the near-surface He clusters mentioned in last paragraph, certain types of damage to the structure of the tungsten substrate are introduced, accompanied by an increase in the He fluence and by nucleation of the He atoms.

*3.2.1 Crowdion bundle in the <111> direction*

A number of studies reported that the formation of He clusters can induce metal atoms being displaced from lattices and becoming self-interstitial atoms (SIAs) [20-21]. The present simulations of cumulative bombardments of He on W provide a more comprehensive picture for the evolution of the SIAs in tungsten. At the early stage of the cumulative bombardments, separate self-interstitial crowdions (SIC) in the <111> direction were generated after the recombination of the SIAs with their neighbouring W atoms. With the fluence of He increasing, bundles of SICs in the

<111> direction appeared. Fig. 6 shows an example of a bundle of two SICs that appear in the case of $E_i$ = 200 eV and $T_s$ = 1500 K. The SIC bundle marked by the arrow appeared after 114 bombardments of He atoms. It was constrained by the He clusters for a long time and moves to and fro around in the <111> direction, as shown in Fig. 6(a-c). This phenomenon was also observed for other SICs after they were formed. It has been reported that single SIC in W bulk can rotate at a substrate that is at a temperature higher than 500K [22-24]. In the present study, the rotations of SIC bundles were also observed. The rotation of SIC is actually the rearrangement of W atoms. Fig. 6(d) shows that the SIC bundle in Fig. 6(a-c) rotated from a <111> direction to another equivalent <111> direction after 2 ps of the 129$^{th}$ bombardment. With the further increase of the He fluence, larger SIC bundles appeared, as is shown in Fig. 7. Rotations of large SIC bundles were hardly observed. In addition to the change in the sizes of the bundles, the number of and sizes of the He clusters that surround the SIC bundles had also been changed. It is still not clear if the occurrence of the rotation of a SIC bundle was influenced by its size or its neighbouring He clusters.

Although a SIC bundle in the <111> direction could be constrained by He clusters for a long time, once the SIC bundle escapes the constraint, it would diffuse along the <111> directions quickly and finally move onto the top of the surface. The migration energy of a SIC in W has been reported to be 0.05 eV in ref. [25-26] and 0.02 eV in ref. [24]. An example of $E_i$ = 200 eV and $T_s$ = 300 K is shown in Fig. 7. The snapshots are obtained in two directions. In Fig. 7(a-f), a large SIC bundle

(denoted by the arrow) moved toward the surface through He clusters Accompanied by the movement, the pressure inside the He cluster was released, the disordered W atoms around the clusters disappeared (Fig. 7(g, h)), and some stacking W atoms that were generated can finally be seen on the top of the surface (Fig. 7(g)).

We also calculated the coordination number $N_{cor}$ of the W atoms in the substrates shown in Fig. 7. The coordination number was calculated by defining the bond-length as 0.9 the lattice length of the W crystal. The results are illustrated in Fig. 8, in which a W atom is represented by a circle if its $N_{cor} > 8$; otherwise, it is represented by a dot. Obviously, combined with the substrate snapshots in Fig. 7, the $N_{cor}$ of most W atoms around He atoms and clusters is greater than 8. Especially in Fig. 8(a, b), a bundle of <111> SICs was hindered and constrained by He clusters; thus, we can see a large cluster (marked by a circle) of W atoms with $N_{cor} > 8$. Once the SIC bundle moved away from the He clusters, the cluster of W atoms disappeared as their $N_{cor}$ became less than or equal to 8 (Fig. 8(c, d)). In the migration path of the SIC bundle, some W atoms that it passed through might increase their $N_{cor}$, as shown in Fig. 8(e, f). When the SIC bundle disappeared, their $N_{cor}$ became less than 9 again (Fig. 8(g, h)). Corresponding to Fig. 7, Fig. 9 shows the number of W atoms with $N_{cor} > 8$ against the number of bombardments. When the SIC bundles were hindered and constrained by the He clusters (before Fig. 7(c)), the number of W atoms with $N_{cor} > 8$ was high, fluctuating between 300 and 360. Then, with the SIC bundles moving to and absorbed on the surface, the number decreased to the neighbourhood of 200.

A bundle of <111> SICs can be regarded as nuclei of the <111> dislocation loop.

Thus, the one-dimensional (1D) diffusion of a bundle of SICs corresponds to the 1D diffusion of dislocation loops. The 1D diffusion of dislocation loops has been experimentally observed in iron. In ref. [27], using transmission electron microscopy, it was found that the nanometer-sized loops with Burgers vector of 1/2 <111> in α-Fe can undergo 1D diffusion. In ref. [28], a similar phenomenon was reported. In face-centred-cubic (FCC) metals after a noble-gas ion (0.5-2 keV) bombardment, a prismatic dislocation loop was formed in the (111) plane, which was finally emitted toward the surface to make an adatom island eruption (AIE) due to the accumulation of local pressure in the primary damaged region inside the substrate. The incident energy of a noble-gas ion in ref. [28] is much higher than that of incident He in our work, thus, one ion can generate two or three AIEs. However, in our work, more He atoms with low energy are required to induce one AIE.

*3.2.2 Dislocation and stacking fault*

In addition to the crowdion bundle in the <111> direction, defects of other types were also observed around the He clusters, as shown in Fig. 10(a). Large He bubbles produce many disordered W atoms on the surface side. By rotating the slice 'A' around the <001> direction by 20 degrees, a staggered dislocation defect can be observed (Fig. 10(b)). We observed slices in parallel with slice 'A' and found more dislocations of the same position, which tells us that a stacking fault was generated. The defects finally evolved to bundles of <111> SICs, which were emitted onto the top surface, as shown in Fig. 10(c, e). With the interstitial defects diffusing away (Fig. 10(d, f)), the pressure inside the He bubbles was released, which promoted the coalescence of He bubbles, as can be seen in Fig. 10(e).

As was shown above, the ruptures of He bubbles and the emitting of interstitial defects cause stacking W atoms on the top the surface. In the increase of the incident He fluence, an obvious surface growth can be observed. For example, in Fig. 11, the surface of a tungsten substrate at 1500 K after 6000 He bombardments with $E_i$=200 eV made a growth of 1.5 lattice lengths in the normal direction of the surface, and the stacking W atoms all stay in lattice positions. We calculated the ratio $R_s$ of the number of stacking W atoms to the number of in-substrate He atoms. An example of $E_i$=200 eV and $T_s$=1500 K is shown in Fig. 12. The data was obtained per 500 He bombardments. In the initial stage of the cumulative bombardments, stacking W atoms are generated mainly due to the primary knock-on of backscattered He atoms and the replacement sequence along the <111> direction [11], and the escaping rate of He atoms is relatively high, as shown in Fig. 4. The value of $R_s$ is thus high. With the He fluence increasing, as we mentioned in section 3.1.2, the temperature of 1500 K accelerates the migration of He atoms and promotes the formation of He clusters, and the escaping rate of the He atoms decreases. The migration of <111> SICs produced by He clusters to the substrate surface now plays the main role of generating stacking W atoms. The value of $R_s$ after 6000 He bombardments is slightly higher than 2. This value is close to the He/vacancy ratio obtained by Wang et al for He in titanium [29] and that reviewed by Wolfer for He in other metals [30].

## 4. Concluding remarks

We presented a MD simulation study on the cumulative bombardments of low-energy (≤200 eV) He atoms on tungsten surfaces. The behaviour of He and the response of the surface were investigated. It is observed that the He incident energy and tungsten temperature have an important influence on the formation and growth of helium clusters. At a high surface temperature, He atoms and clusters are easier to combine and coalesce into large bubbles, which will hinder both the escape and penetration of the He atoms.

During the formation of the helium clusters, interstitial W atoms are produced. They are disordered initially and can produce interstitial defects, which can evolve into bundles of <111> crowdions and would be constrained around the He clusters for a long time. Once they are free of the constraints, they will finally move onto the top of the surfaces along the <111> direction, resulting in stacking W atoms on the surface. The process above appears frequently in the whole simulation.

Instantaneous rupture of the helium clusters has also been observed. With the decline in the cluster pressure, some surface W atoms nearby gradually moved downward and resulted in the healing of the pore that was left. In addition, a transient disorder of the tungsten surface structure occurred occasionally, accompanying the escape and re-deposition of the W atoms.

Although the fluence of incident He that was achieved in our simulations was still much smaller than the fluence that can be achieved experimentally, it has been shown that the dynamical processes that occurred during the cumulative bombardments are much more complex than what occurs in cases in which the incident fluence is sparse. To perform molecular dynamics simulations under conditions that are comparable with the experimental conditions would be excessively challenging. Monte Carlo methods and rate-theory-based methods are more computationally efficient. However, it is likely that the physics models in these methods must be reconsidered for the cases of high-incident fluence, where according to what is depicted in the present study, the interactions among multiple defects (i.e., He clusters with He clusters, He clusters with interstitial defects and interstitial defects with interstitial defects) play important roles on the structure evolution of the substrates. In addition, the physics models for the release of He atoms and the accompanying ejection of W atoms must also be reconsidered given that the production rates of various impurities are always matters of concern in fusion plasmas.

**Acknowledgments**

This work was partially supported by the National Natural Science Foundation of China (Contact Nos. 91126001 and 11175124) and the National Magnetic Confinement Fusion Program of China (2013GB109002 and 2011GB110005).

**Figure Captions**

**Table 1.** The box sizes and the corresponding He fluences that were reached in our simulations for different incident energies $E_i$.

**Fig. 1.** The number of escaping He atoms from tungsten surfaces at 1500 K as a function of the number of injected He atoms. Square: $E_i$=60 eV; circle: $E_i$=100 eV; triangle-up: $E_i$=200 eV.

**Fig. 2.** The tungsten substrates with the same retention dose of He atoms. $T_s$=1500 K. (a) $E_i$=100 eV; (b) $E_i$=200 eV. Gray dots: W atoms; dark balls: He atoms.

**Fig. 3.** The tungsten substrates with the same number of retained He atoms. $E_i$=200 eV. (a) $T_s$=300 K; (b) $T_s$=1500 K.

**Fig. 4.** The number of escaping He atoms per 50 injected He atoms as a function of the number of injected He atoms. $E_i$=200 eV. Square: $T_s$=300 K; circle: $T_s$=1500 K.

**Fig. 5.** The rupture process of the He cluster 'A' and the accompanying transient disorder of the surface structure, shown along two directions. $E_i$=100 eV, $T_s$=1500 K. Left column: <100>; right column: <001>. (a) and (b) t=0.00 ps; (c) and (d) t=1.52 ps; (e) and (f) t=2.20 ps; (g) and (h) t=6.00 ps.

**Fig. 6.** The migration and rotation of 2 <111> SICs. $E_i$=200 eV, $T_s$=1500 K. The $121^{th}$ ((a) t=2.00 ps; (b) t=4.00 ps; (c) t=6.00 ps) and $129^{th}$ ((d) t=2.00 ps) He atom bombardment.

**Fig. 7.** The behaviour of a bundle of SICs shown along two directions. $E_i$=200 eV, $T_s$=300 K. Left column: <010>; right column: <001>. Snapshots that were all obtained at t=10.00 ps of the $1966^{th}$ ((a), (b)), $1967^{th}$ ((c), (d)), $1968^{th}$ ((e), (f)) and

1969$^{th}$ ((g), (h)) He bombardment.

**Fig. 8.** $N_{cor}$ of the W atoms in the substrates shown in Fig. 7. $E_i$=200 eV, $T_s$=300 K. Left column: <010>; right column: <001>. Gray dots: W atoms with $N_{cor} \leq 8$; circles: W atoms with $N_{cor} > 8$.

**Fig. 9.** The number of W atoms with $N_{cor} > 8$ as a function of the amount of He bombardment. $E_i$=200 eV, $T_s$=300 K.

**Fig. 10.** The behaviour of a complicated defect. $E_i$=200 eV, $T_s$=1500 K. The 3070$^{th}$ He bombardment. Left column: substrates; right column: slice 'A' after a 20 degrees rotation around the <001> direction. (a) and (b) t=0.6 ps; (c) and (d) t=2.00 ps; (e) and (f) t=10.0 ps.

**Fig. 11.** Surface growth. $E_i$=200 eV, $T_s$=1500 K. Left: initial substrate; right: substrate after 6000 He bombardments.

**Fig. 12.** The ratio $R_s$ as a function of the amount of He bombardment. $E_i$=200 eV, $T_s$=1500 K.

**Table 1.**

| $E_i$ | $T_s$ | Box size (Å$^3$) | Fluence (10$^{20}$ m$^{-2}$) |
|---|---|---|---|
| 60 eV | 1500 K/2100K | 63.3×63.3×63.3 | 0.50 |
| 100 eV | 1500 K/2100K | 63.3×63.3×94.9 | 2.50 |
| 200 eV | 300 K/1500K | 94.9×94.9×94.9 | 1.11 |

**Fig. 1.**

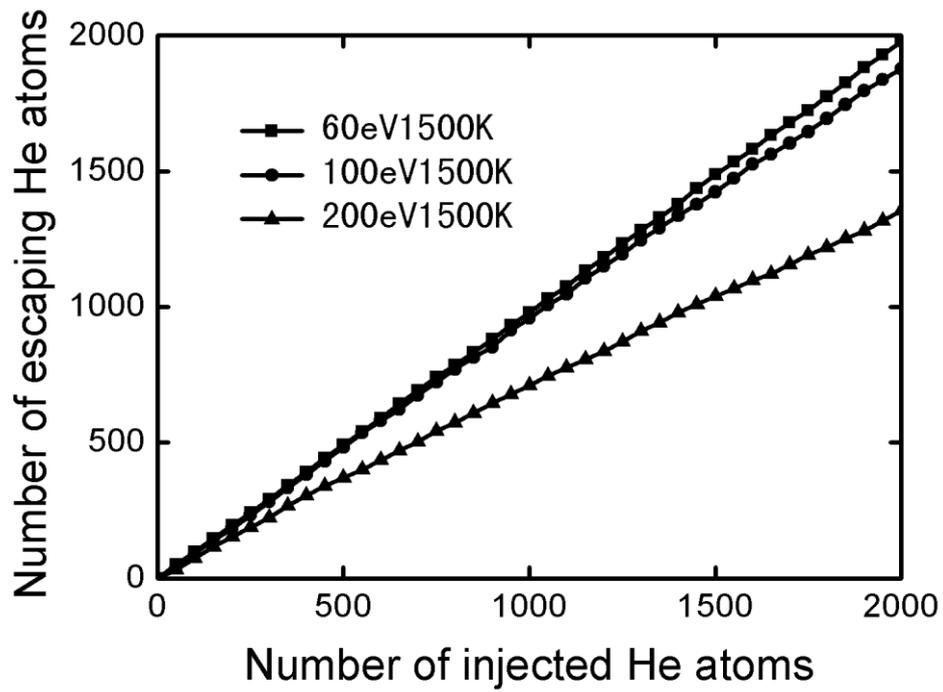

**Fig. 2.**

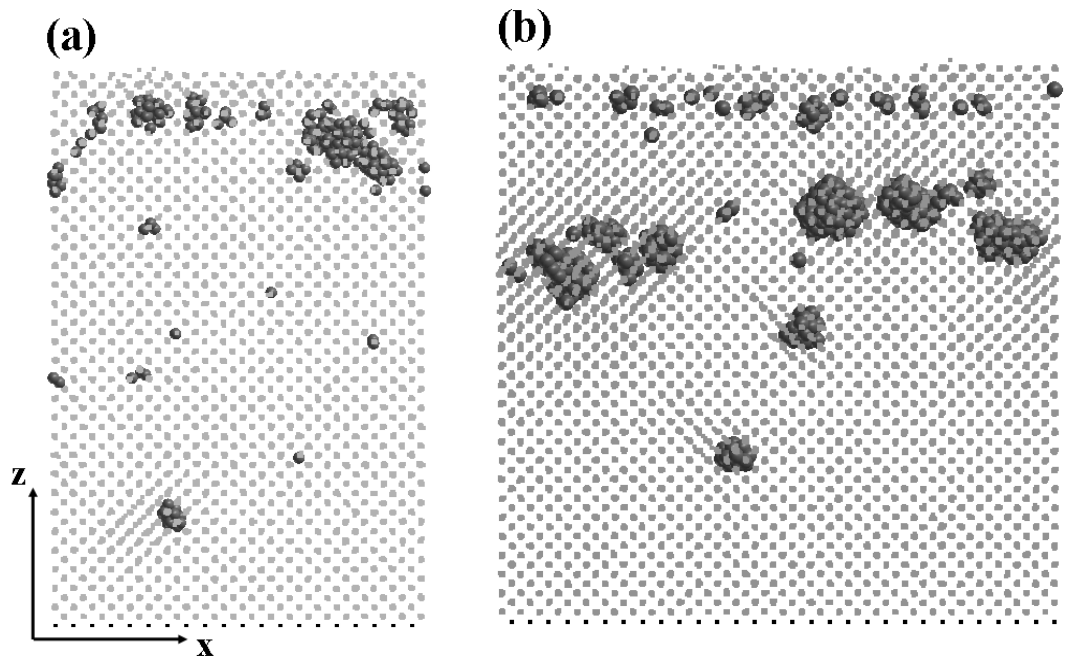

**Fig. 3.**

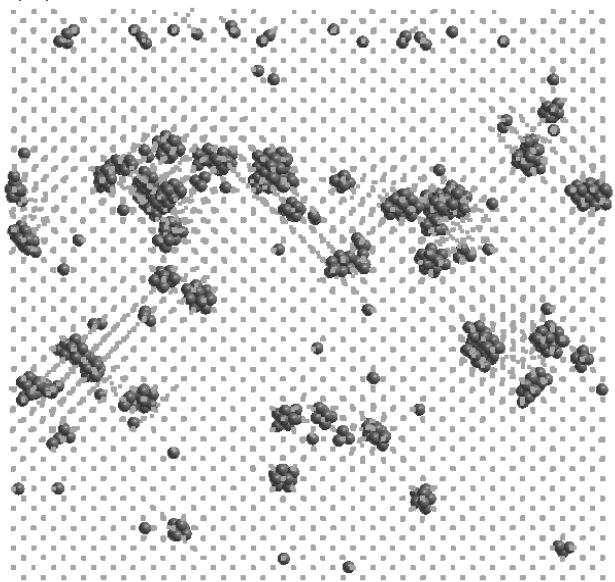

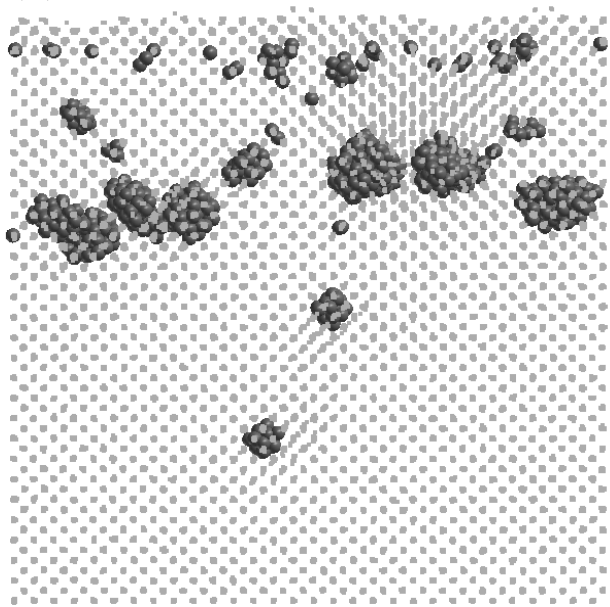

Fig. 4.

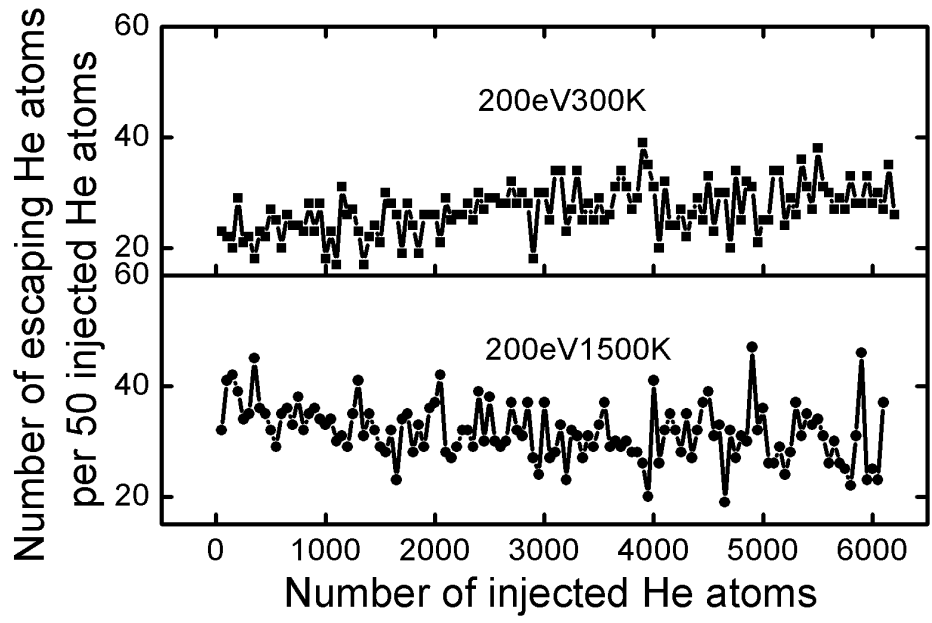

**Fig. 5.**

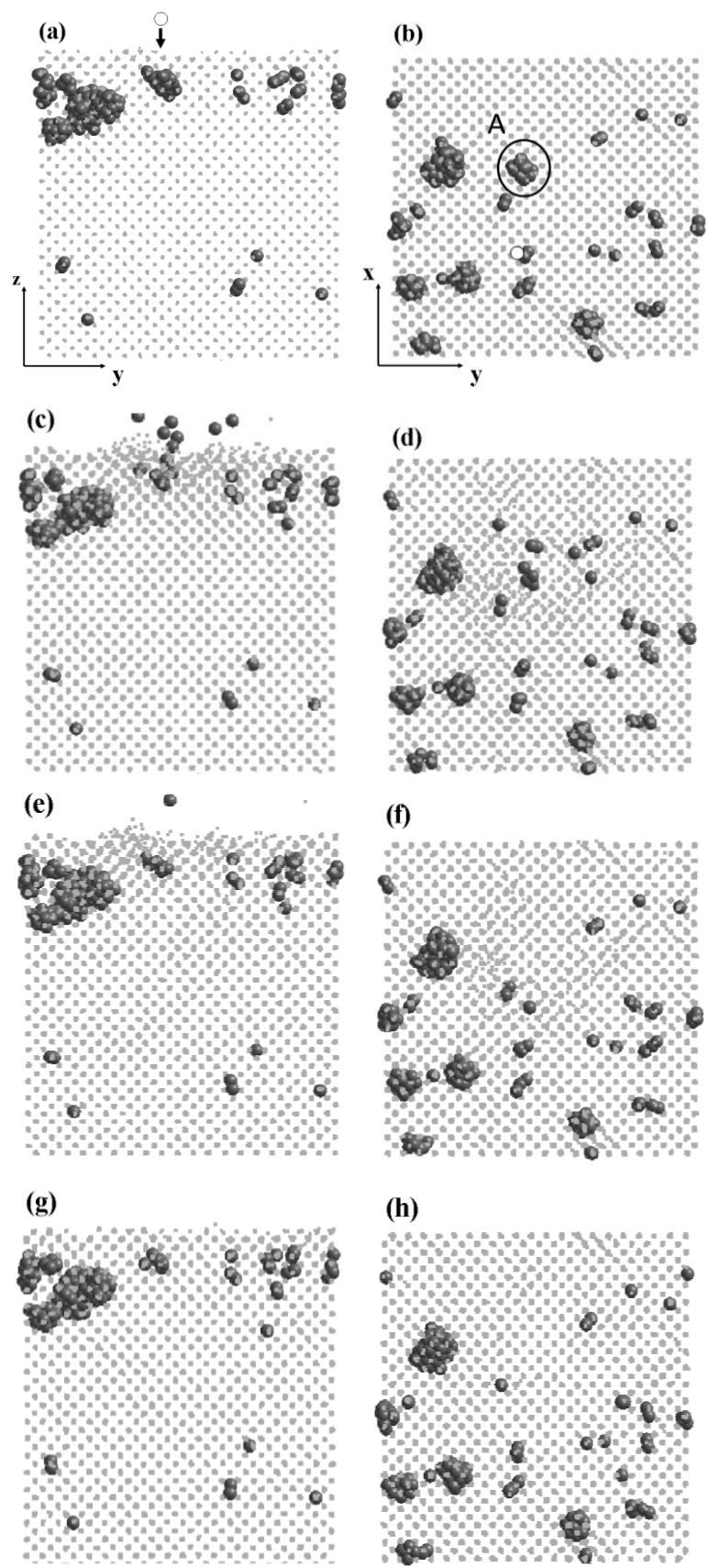

**Fig. 6.**

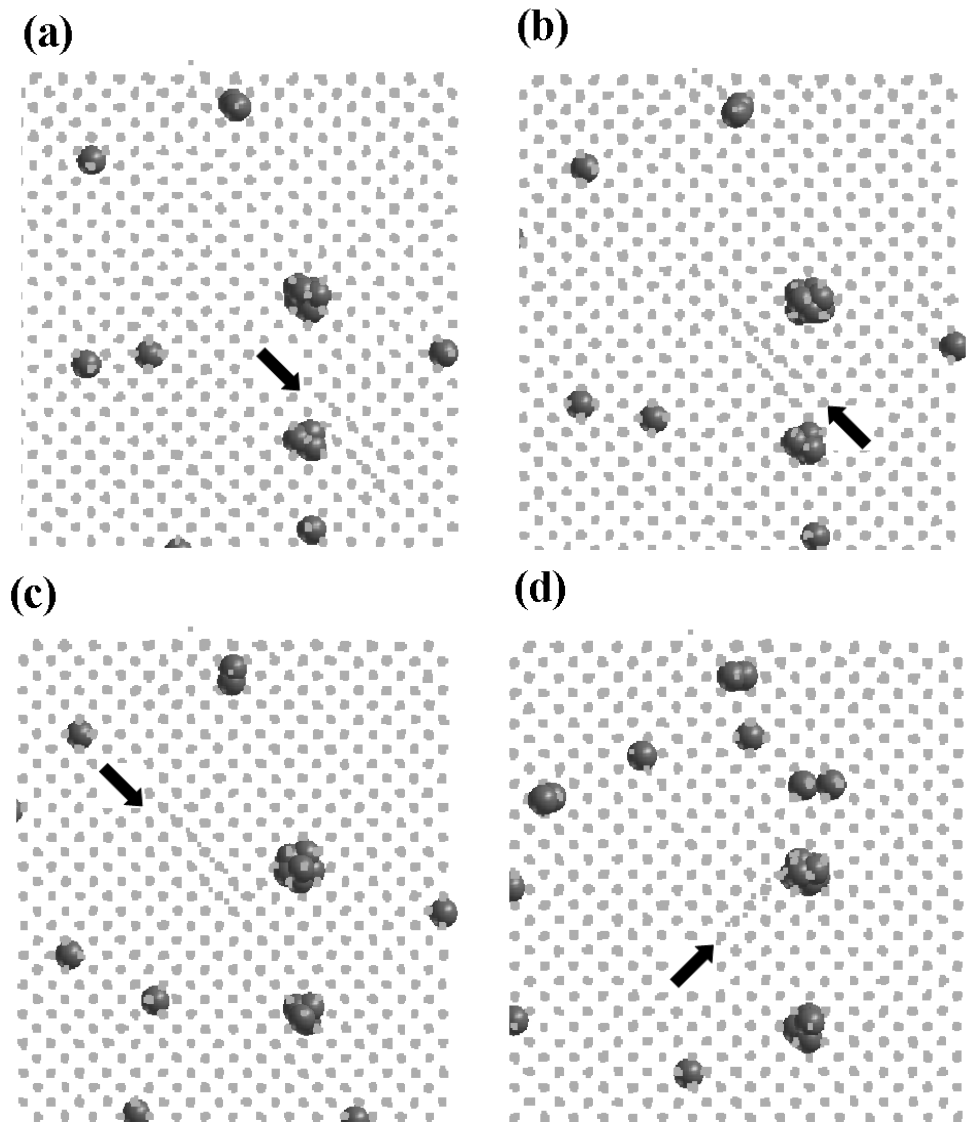

**Fig. 7.**

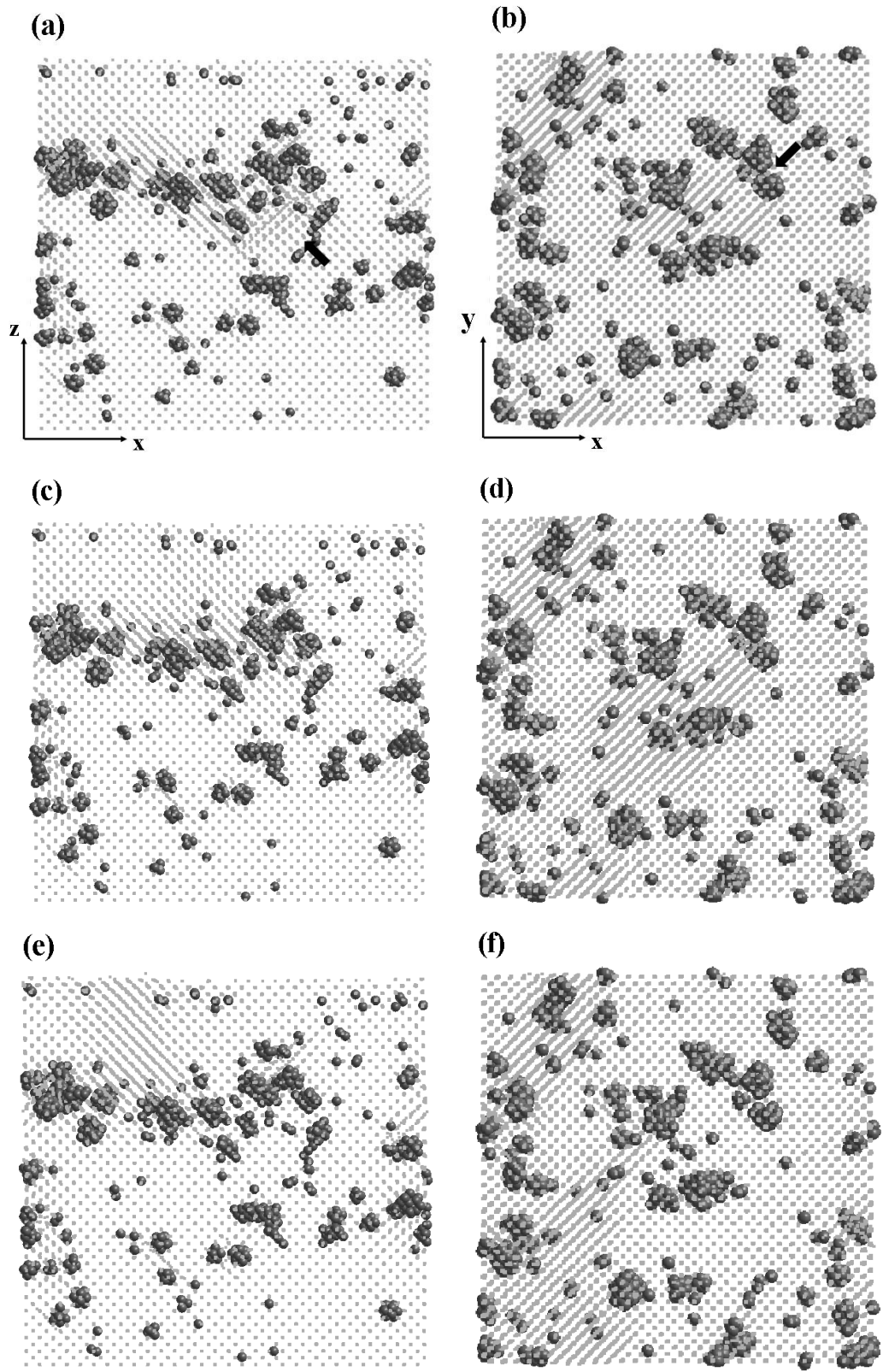

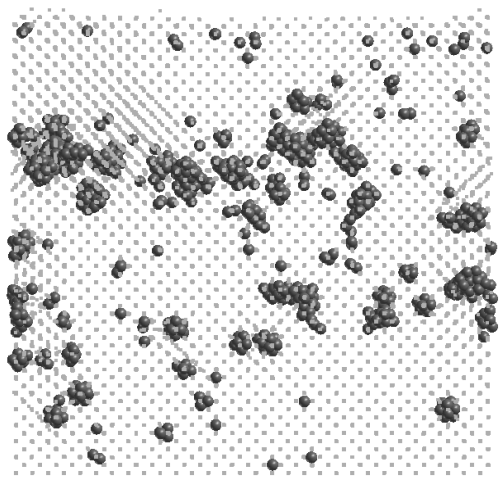 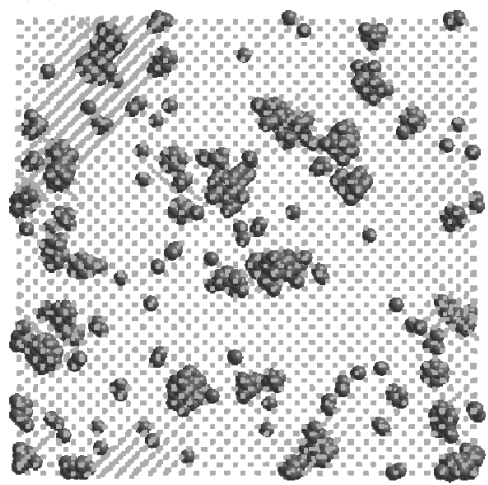

**Fig. 8.**

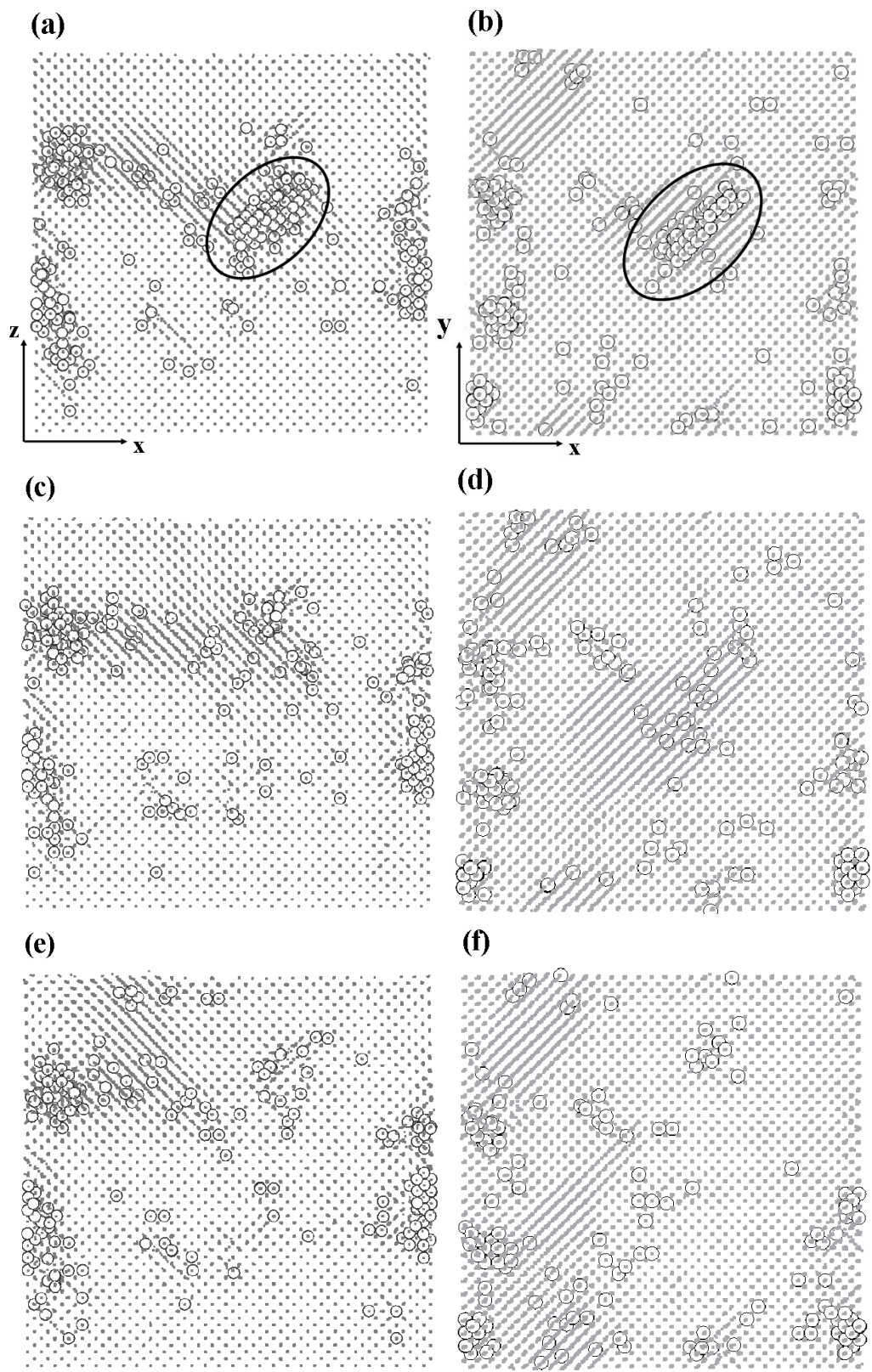

**(g)** 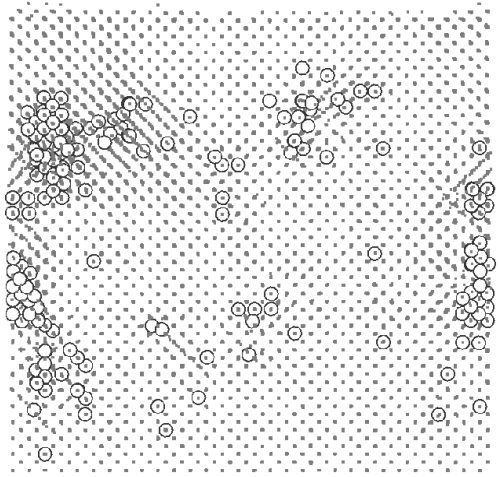 **(h)** 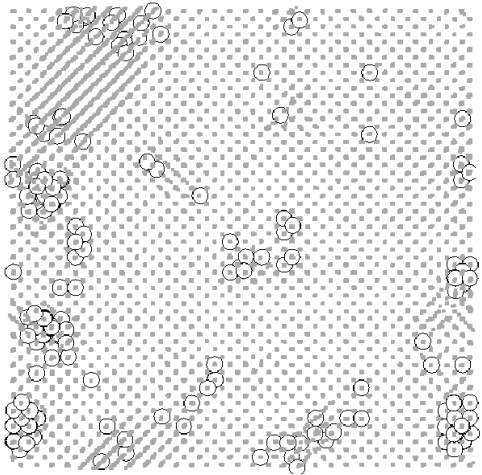

Fig. 9.

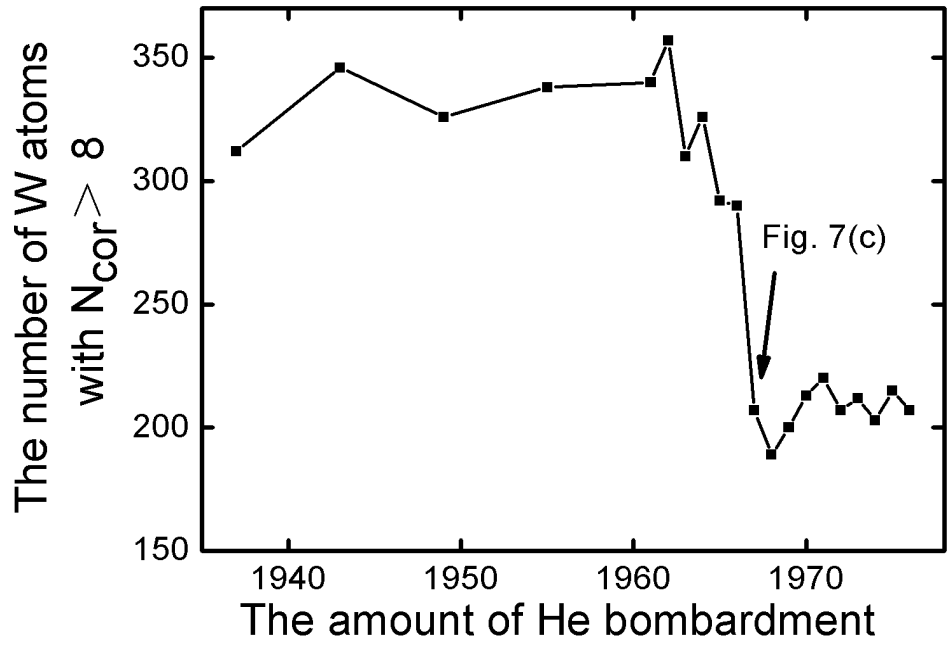

Fig. 7(c)

**Fig. 10.**

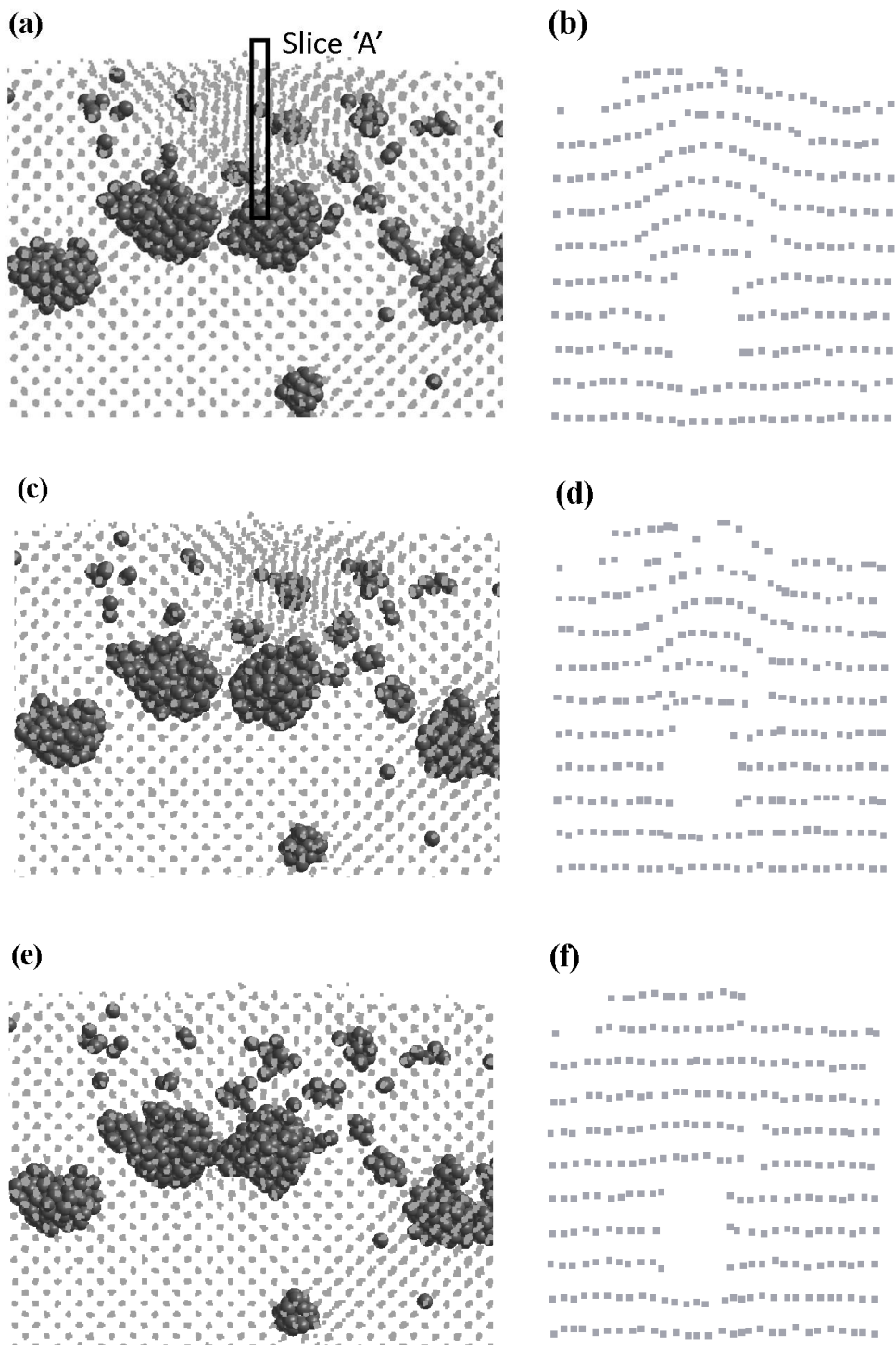

**Fig. 11.**

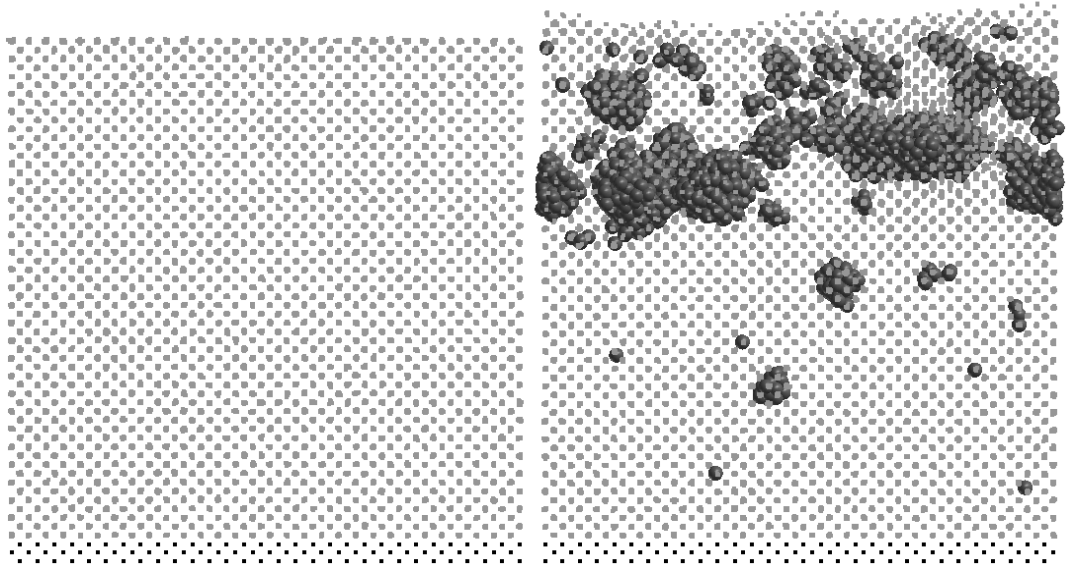

**Fig. 12.**

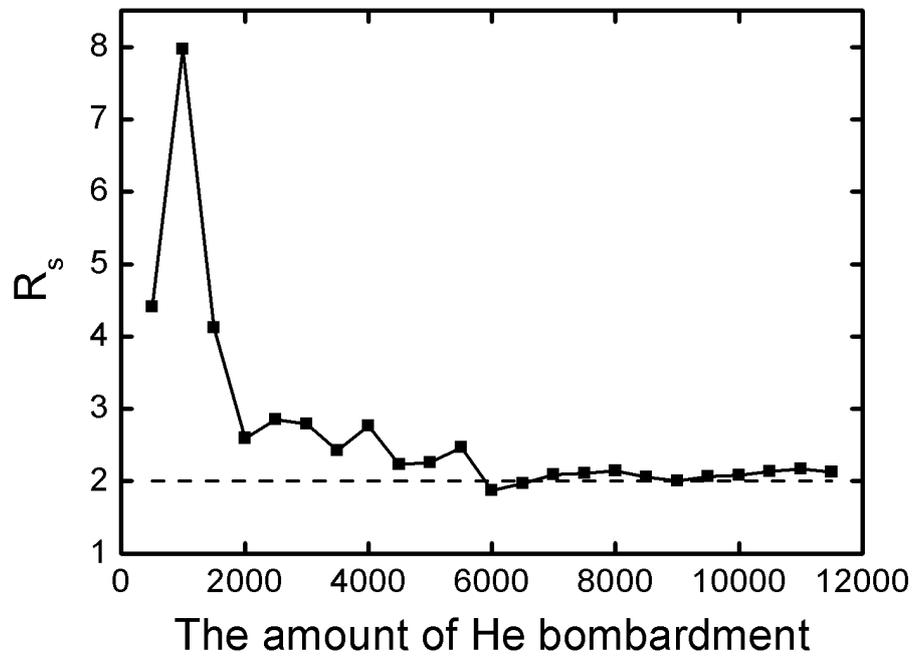